# Evolution of Spin Relaxation Processes in LiY$_{1-x}$Ho$_x$F$_4$ with Increasing $x$ Studied via AC-Susceptibility and Muon Spin Relaxation

R. C. Johnson,[1] B. Z. Malkin,[2] J. S. Lord,[3] S. R. Giblin,[3] A. Amato,[4] C. Baines,[4] A. Lascialfari,[5] B. Barbara,[6] and M. J. Graf [1,*]

[1] Department of Physics, Boston College, Chestnut Hill, MA 02467 USA
[2] Kazan Federal University, Kazan 420008, Russian Federation
[3] Rutherford Appleton Laboratory, Didcot, Oxfordshire OX11 0QX, UK
[4] Paul Scherrer Institute, CH 5232 Villigen PSI, Switzerland
[5] Inst. of General Physiology and Biological Chemistry, Univ. of Milan, I20134 Milan, Italy
[6] Néel Institute, Dept. of Nanosciences, CNRS, 38042 Grenoble Cedex-09, France

We present measurements of magnetic field and frequency dependences of the low temperature ($T$ = 1.8 K) AC-susceptibility, and temperature and field dependences of the longitudinal field positive muon spin relaxation (μSR) for LiY$_{1-x}$Ho$_x$F$_4$ with $x$ = 0.0017, 0.0085, 0.0408, and 0.0855. The fits of numerical simulations to the susceptibility data for the $x$ = 0.0017, 0.0085 and 0.0408 show that Ho-Ho cross-relaxation processes become more important at higher concentrations, signaling the crossover from single-ion to correlated behavior. We simulate the muon spin depolarization using the parameters extracted from the susceptibility, and the simulations agree well with our data for samples with $x$ = 0.0017 and 0.0085. The μSR data for samples with $x$ = 0.0408 and 0.0855 at low temperatures ($T$ < 10 K) cannot be described within a single-ion picture of magnetic field fluctuations and give evidence for additional mechanisms of depolarization due to Ho$^{3+}$ correlations. We also observe an unusual peak in the magnetic field dependence of the muon relaxation rate in the temperature interval 10 – 20 K that we ascribe to a modification of the Ho$^{3+}$ fluctuation rate due to a field induced shift of the energy gap between the ground and the first excited doublet crystal field states relative to a peak in the phonon density of states centered near 63 cm$^{-1}$.

# I. INTRODUCTION

Over the past forty years LiY$_{1-x}$Ho$_x$F$_4$ has proven to be a nearly ideal system for studying magnetic phenomena.[1] LiHoF$_4$ is a textbook example of a dipolar Ising ferromagnet, with $T_C$ = 1.51 K,[2] and dilution of the magnetic Ho$^{3+}$ ions with non-magnetic Y$^{3+}$ ions yields a frustrated ferromagnet which in turn gives way to a spin glass phase for $0.05 < x < 0.25$.[3,4] The lattice constants are almost unchanged with a variation of $x$, and large, high quality single crystals are available, making this system ideal for both experimental and theoretical studies. The very dilute limit ($x < 0.005$) has been recently studied as well, and is a paradigm system for studying quantum tunneling of the magnetization;[5] for example, the single-ion spin dynamics have been accurately modeled to reproduce AC-susceptibility[6,7] and $^{19}$F nuclear spin-lattice relaxation data.[8] However, the nature of the low-temperature ground state for intermediate concentrations remains controversial. Indeed, three recent studies of the AC-susceptibility at very low temperatures yield very different conclusions: no conventional spin glass,[9] conventional spin glass[10,11] and quantum spin glass[4] phase transition. In part, interpretation of the data is difficult due to the very long relaxation times at low temperatures ($T < 100$ mK).[12]

The LiY$_{1-x}$Ho$_x$F$_4$ crystals have a body centered tetragonal Scheelite structure with $C_{4h}$ space group and $S_4$ point symmetry group at Ho$^{3+}$ sites (for $x = 0$, the lattice constants are $a = b = 5.175$Å, $c = 10.74$ Å).[13] In the dilute limit the physics is controlled by the single-ion Ho$^{3+}$ energy spectrum. The ground $^5I_8$ multiplet of the electronic $4f^{10}$ configuration is split by the crystal field into a ground state Ising doublet separated from the first excited singlet state by a gap of 9.8 K; hyperfine coupling with the $I = 7/2$ $^{165}$Ho (100% abundance) nucleus produces a manifold of states consisting of 8 doubly-degenerate levels separated by roughly 200 mK energy gaps. Application of a magnetic field along the quantization axis (crystalline $c$-axis) induces a sequence of avoided level crossings (ALCs) at resonant fields given approximately by $B_n = n \times 23$ mT ($-7 \leq n \leq 7$), at which enhanced spin relaxation rates are observed.[5]

It was realized earlier that the observed field-dependent AC-susceptibility of low concentration samples (of the order of 0.1 at. % Holmium) must be described in terms of multi-spin relaxation processes.[14] This was further developed into a microscopic theory of relaxation rates through numerical simulations which take into account the effects of crystal field, electron-phonon, and hyperfine interactions, as well as cross-relaxation processes.[6,7] At low frequencies, peaks (dips) in the field dependences of the in-phase and out-of-phase susceptibilities (hereafter referred to as $\chi'$ and $\chi''$, respectively) indicate enhanced relaxation processes that occur at the field-induced ALCs. Additional peaks (dips) in $\chi'$ ($\chi''$) at $n > 7$, along with smaller peaks in $\chi'$ and $\chi''$ at half-integer $n$ values (-13 ≤ 2$n$ ≤ 13), have also been observed and explained. Even at very low concentrations, it was critical to include in the model both multi-spin and phonon bottleneck effects in order to qualitatively describe these results. Although there are many different relaxation processes occurring within the sample, an effective relaxation rate of the magnetization at ALCs can be identified by a 'dip-to-peak' crossover in $\chi''$ with increasing frequency. Also, the frequency dependence of the dynamic susceptibility shows a roll-off in $\chi'$ and a maximum in $\chi''$ centered about a characteristic frequency $f_0$ of $Ho^{3+}$ spin fluctuations. We look for the above signatures in our low concentration samples, and expect both the effective relaxation rate and $f_0$ to increase with concentration.[5,6]

In this work, we report AC magnetic susceptibility and longitudinal field μSR measurements on four single crystal samples of $LiY_{1-x}Ho_xF_4$ with holmium concentrations $x$ in the range 0.0017 - 0.085. Our goal is to pursue a bottom-up approach to understanding correlations in this system and the transformation of single-ion dynamics into collective dynamics. We find that both the susceptibility and μSR data for our sample with intermediate concentration can be numerically simulated within a single-ion model modified to include Ho-Ho cross-relaxation, but with parameter values that signal the crossover to correlated behavior. Samples at higher concentration show qualitatively different behavior outside the scope of this model. Finally we report an unusual peak in the field-dependent muon depolarization rate at intermediate temperatures, which we ascribe to a modification of the $Ho^{3+}$ fluctuation rate due to a field induced shift of the energy gap between the ground and the first excited doublet crystal field sublevels of the $^5I_8$ multiplet relative to a peak in the phonon density of states centered at 63 cm$^{-1}$. This interpretation is supported by numerical simulations of the muon depolarization which qualitatively reproduce the peak.

## II. EXPERIMENTAL

High quality single crystal samples of $LiY_{1-x}Ho_xF_4$ with nominal holmium concentrations of 0.2, 1, 5, and 10 at. % were determined to have concentrations of $x$ = 0.0017, 0.0085, 0.0408 and 0.0855 via the field dependences of the magnetization in units of [emu/per sample gram] measured at $T = 2$ K in the DC- fields from 0 to 3 T parallel to the sample $c$- axis. These data were compared with the calculated field dependence of the magnetization in units of [emu/per $Ho^{3+}$ mol], and the latter curve was scaled until it matched the measured one, thereby determining the concentration. For AC-susceptibility measurements the samples were cut to dimensions of approximately $2\times2\times8$ mm$^3$ in order to minimize demagnetization effects, with the c-axis aligned along the 8 mm edge and mounted parallel to the applied DC field. Accordingly, demagnetizing fields have been neglected in the simulations described below in Section III. By comparing the field dependence of the theoretically predicted locations of the ALCs with our measurements for the $x = 0.0017$ sample, we determined that our alignment of the $c$-axis is approximately $6.0 \pm 2.4$ degrees off of the applied field. It should be noted that, according to X-ray diffraction measurements on three pieces of $x = 0.0085$ sample, the crystal has been in fact cut between 6.3 and 6.8 degrees from the $c$-axis, indicating a small systematic misorientation. The susceptibility simulations presented below account for the small measured misalignments, although the field axes in the figures present the absolute value of the applied field.

We probed the bulk magnetic response using two different susceptometers to measure AC-susceptibility as a function of magnetic field in a gas flow cryostat ($T \geq 1.7$ K) with a superconducting magnet (0 to 9 T). The first was the MagLab system manufactured by Oxford Instruments with a frequency range up to 10 kHz. This susceptometer was calibrated with measurements of a superconducting Nb sphere to confirm that the in- and out-of-phase components of the signal correctly corresponded respectively to the real and imaginary parts of the susceptibility $\chi = \chi' + i\chi''$. The second susceptometer was homebuilt with hand wrapped coils and a frequency range up to 100 kHz. A calibration for this susceptometer was obtained through a comparison of identical experiments done in both systems. Measurements of AC-susceptibility were carried out by driving the system with a sinusoidally varying field parallel the DC field and crystal $c$-axis, with an amplitude 0.4 mT and 0.01 mT on the MagLab and homebuilt susceptometers, respectively. Excitation amplitudes and frequencies were chosen to prevent unwanted heating in the sample and susceptometer and are consistent with values used in previous work.[14] Relatively slow DC field sweep rates of 0.02 or 0.24 T/min were used, depending on the density of points desired, so that we could observe relaxation of the magnetization at quasi-static fields.[5] Data for both rates were compared and no difference outside of error was seen.

To complement the bulk susceptibility measurements, positive muon spin relaxation (μSR) measurements were made to probe the local magnetic properties. In time-differential μSR, spin polarized positive muons are implanted in the sample, and the time-dependent response of the muon spin orientation to the local magnetic environment is monitored via the positrons emitted from millions of muon decay events. From this one extracts the asymmetry $A(t)$ (the ratio of the difference to the sum of the decay positrons recorded by a set of counters placed forward and backward of the sample with the respect to the initial polarization of the muon beam), which is proportional to the muon spin autocorrelation function (or depolarization function) $G(t)$. The time dependence of $G$ results from quasi-static (for example, the spatial disorder present in the local magnetic field) and dynamical (such as local field fluctuations) processes. A description of the technique is given in Refs. 15-17. Experiments were run at the pulsed muon source ISIS and the continuous muon source at the Paul Scherrer Institute. The ISIS data were taken in conventional polarization mode, where the muons are 100% polarized with the spin directed anti-parallel to the beam momentum. The PSI data were taken in spin rotated mode, where the muon spin is oriented at approximately 50° to the beam momentum, allowing for simultaneous measurement of the depolarization for spin components in directions parallel and perpendicular to the beam momentum. Data at both facilities were taken over the temperature range 1.8 K < $T$ < 50 K in gas flow cryostats. Samples cut to approximately 2×8×30 mm³ were mounted with the $c$-axis either parallel or perpendicular to the applied field. The muons are known to stop between fluorine sites and form the F⁻-μ⁺-F⁻ bound state, but the coherent spin oscillations of this three-spin system are suppressed for the range of longitudinal fields studied in this work.[18]

These samples have been previously studied by ¹⁹F nuclear magnetic resonance[8,19] and transverse field and zero field muon spin relaxation (μSR).[18,20,21]

## III. SIMULATIONS
### A. AC-Susceptibility

First, we briefly review the model used to simulate the AC-susceptibility, as recounted in detail in Ref. 6. Energies $E_n$ ($n$ = 1 to 136) and the corresponding wave functions of electron-nuclear sublevels of the ground multiplet ⁵I₈ of the ¹⁶⁵Ho³⁺ ions with the nuclear spin $I$ = 7/2 were obtained by numerical diagonalization of the Hamiltonian

$$H_0 = H_{CF} + H_{hf} + H_Q + H_Z + H_{RCF}, \qquad (1)$$

containing the crystal field energy ($H_{CF}$), the hyperfine magnetic ($H_{hf}=A\mathbf{J}\cdot\mathbf{I}$) and quadrupole ($H_Q$) interactions, the electronic Zeeman energy $H_Z = -\mathbf{m}\cdot\mathbf{B}$ ($\mathbf{m} = -g_J\mu_B\mathbf{J}$ is the electronic magnetic moment, $g_J$ is the Landé factor, $\mu_B$ is the Bohr magneton, $\mathbf{J}$ is the total electronic angular moment, $\mathbf{B}$ is the external magnetic field), and the effective interaction with random crystalline strains $H_{RCF}$.

The dynamic susceptibility of a highly dilute sample at temperature $T$ is considered as the sum of single-ion susceptibilities[6]

$$\chi_{\alpha\beta}(\omega) = \chi_{\alpha\beta}^0 - i\omega \sum_{nk} \Delta m_{\alpha,nn}(i\omega\mathbf{1}+W)^{-1}_{nk}\Delta m_{\beta,kk}\rho_{0k}/k_BT$$
$$+ \sum_{n,k\neq n} m_{\alpha,nk}m_{\beta,kn}(\rho_{0k}-\rho_{0n})\left[\frac{1}{\text{h}(\omega_{nk}-\omega-i\gamma_{nk})} - \frac{1}{\text{h}\omega_{nk}}\right], \quad (2)$$

where the static susceptibility equals

$$\chi_{\alpha\beta}^0 = \sum_n \Delta m_{\alpha,nn}\Delta m_{\beta,nn}\rho_{0n}/k_BT + \sum_{n,k\neq n}\frac{m_{\alpha,nk}m_{\beta,kn}}{\text{h}\omega_{nk}}(\rho_{0k}-\rho_{0n}). \quad (3)$$

Here $\gamma_{nk}$ are the homogeneous widths of transitions with frequencies $\omega_{nk}=(E_n-E_k)/\text{h}$, $k_B$ is the Boltzmann constant, $\rho_{0n} = \exp(-E_n/k_BT)/\sum_p \exp(-E_p/k_BT)$ is the relative population of the state $n$, $m_{\alpha,nk}$ and $\Delta m_{\alpha,nn}$ are matrix elements of the electronic magnetic moment and its fluctuation ($\Delta \mathbf{m} = \mathbf{m} - <\mathbf{m}>$) components, respectively, and $W$ is the relaxation matrix with the components

$$W_{nm} = W_{m\rightarrow n}^{(r)} + \sum_{lp}(W_{np,lm}^{CR}\rho_{0p} + W_{nm,lp}^{CR}\rho_{0p} - W_{pn,lm}^{CR}\rho_{0n}) \quad (n\neq m), \quad (4)$$

$$W_{nn} = -\sum_{m\neq n} W_{mn}.$$

The first term on the right hand side of Eq. (4) corresponds to energy exchange between the Ho$^{3+}$ ion and the phonon bath. The renormalization of the one-phonon transition probabilities $W_{m\rightarrow n} = w_{mn}|e^{\text{h}\omega_{mn}/k_BT}(e^{\text{h}\omega_{mn}/k_BT}-1)^{-1}|$ due to the phonon bottleneck effect has been taken into account in the simulations via

$$W_{m\rightarrow n}^{(r)} = W_{m\rightarrow n}\left[1+\frac{P_{mn}w_{mn}|\rho_{0m}-\rho_{0n}|}{\omega_{mn}^2[1+i\omega\tau_{ph}(\omega_{mn})]}\right]^{-1} \quad (5)$$

where $\tau_{ph}(\omega_{mn})$ is the lifetime of resonant phonons, and $w_{mn}$ is the rate of spontaneous phonon emission. Two different factors $P_{mn} = 2\pi^2 v^3 \tau_{ph}(\omega_{mn})N/3\Delta\omega_{mn}$ ($v$ is the average sound velocity, $N$ is the number of the $Ho^{3+}$ ions per unit volume, and $\Delta\omega_{mn}$ is the total width of the transition $m$-$n$) for the low-frequency transitions between the hyperfine sublevels of the ground electronic doublet ($P_g$) and for the high-frequency transitions between the hyperfine sublevels of the first excited singlet $\Gamma_2^1$ and the ground doublet ($P_e$) were considered as fitting parameters, which we report in Table I and discuss in the next section. Note that decreases of $P_e$ and $P_g$ indicate an increase of phonon relaxation rates.

The second term in Eq. (4) accounts for Ho-Ho cross relaxation processes. The rates were calculated according to the expression

$$W^{CR}_{np,lm} = (\delta/h)^2 g^{CR}(\omega_{pn} - \omega_{lm})\{k_{11}(|J_{x,np}J_{x,lm}|^2 + |J_{y,np}J_{y,lm}|^2) \\ + k_{12}(J_{x,np}J_{x,lm}J_{y,pn}J_{y,ml} + J_{y,np}J_{y,lm}J_{x,pn}J_{x,ml}) + k_{33}|J_{z,np}J_{z,lm}|^2 \\ + k_{13}[J_{z,np}J_{z,lm}(J_{x,pn}J_{x,ml} + J_{y,pn}J_{y,ml}) + J_{z,pn}J_{z,ml}(J_{x,np}J_{x,lm} + J_{y,np}J_{y,lm})] \\ + k_{44}(|J_{x,np}J_{z,lm} + J_{z,np}J_{x,lm}|^2 + |J_{y,np}J_{z,lm} + J_{z,np}J_{y,lm}|^2) \\ + k_{66}|J_{x,np}J_{y,lm} + J_{y,np}J_{x,lm}|^2 + \varepsilon^2|C^{(2)}_{2,np}C^{(2)}_{-2,lm} + C^{(2)}_{-2,np}C^{(2)}_{2,lm}|^2\}, \quad (6)$$

where the cross-relaxation form function was considered as Gaussian with dispersion $\Delta$:[22]

$$g^{CR}(\omega_{pn} - \omega_{lm}) = \frac{1}{\sqrt{2\pi}\Delta}\exp[-(\omega_{pn} - \omega_{lm} - \frac{h\Delta^2}{2k_BT})^2/2\Delta^2]. \quad (7)$$

The parameter $\delta$ is the average dipolar interaction energy between two $Ho^{3+}$ ions, and the last term in Eq. (6) has been introduced to account for the interaction between the $Ho^{3+}$ ions through the dynamical lattice deformations.[23] In the simulations, we have accounted for all possible cross-relaxation and one phonon transitions between the lower 64 electron-nuclear states of the $Ho^{3+}$ ions corresponding to the ground crystal field doublet $\Gamma_{34}^1$, excited singlets $\Gamma_2^1$ (6.85 cm$^{-1}$), $\Gamma_2^2$ (23 cm$^{-1}$), $\Gamma_1^1$ (48 cm$^{-1}$), $\Gamma_1^2$ (57 cm$^{-1}$), and the first excited doublet $\Gamma_{34}^2$ (72 cm$^{-1}$). The terms in parentheses indicate corresponding crystal field energies. All concentration-independent parameters which we use in simulations of the AC-susceptibilities have been determined in previous studies of optical and EPR spectra, $^{19}$F nuclear spin relaxation and AC-susceptibilities in low-$x$ LiY$_{1-x}$Ho$_x$F$_4$ samples ($x \leq 0.0027$).[6,7,8,24] In particular, crystal field parameters in the Hamiltonian $H_{CF}$ and electron-phonon coupling constants used in calculations of the relaxation rates are given in Table I in Ref. 6; magnetic and quadrupolar hyperfine coupling constants, and factor Lande ($g_J = 1.21$) are given in Ref. 24; values of the phonon lifetimes $\tau_{ph}$ in Eq. (5) and the dimensionless parameters $k_{ij}$ in Eq. (6) are given in Ref. 8 (p.162). The parameters of the low-symmetry component of the crystal field in the Hamiltonian $H_{RCF}$ which has been introduced to account for the anti-crossings induced by random strains[24] are fixed at the values $B_2^2 = -B_2^{-2} = 1$ cm$^{-1}$ obtained earlier in Ref. 6 and are not varied in the present work.

### B. Muon Spin Relaxation

The probability of a transition per unit time between the two muon spin states $S_z = \pm 1/2$ (the quantization axis is directed along the magnetic field $\boldsymbol{B}$) due to fluctuations of the magnetic moment of the Ho$^{3+}$ ion at a distance $r$ from a muon is given by the expression

$$w(\boldsymbol{r}) = C(\boldsymbol{r}/r, \boldsymbol{B}, T)/2r^6 \qquad (8)$$

where

$$C(\boldsymbol{r}, \boldsymbol{B}, T) = \frac{\gamma^2}{2} \{J_{\Delta m \Delta m}(\omega_r) + 3J_{\Delta m_r \Delta m_r}(\omega_r) - J_{\Delta(m_B - 3m_r e_{rB})\Delta(m_B - 3m_r e_{rB})}(\omega_r)\}, \qquad (9)$$

$\gamma$ is the muon gyromagnetic ratio, $\omega_r = \gamma B_{loc}$ is the muon Larmor frequency in the local magnetic field

$$\boldsymbol{B}_{loc} = \boldsymbol{B} + [-<\boldsymbol{m}> r^2 + 3\boldsymbol{r}(\boldsymbol{r} \cdot <\boldsymbol{m}>)]/r^5, \qquad (10)$$

and $\Delta m_r = \Delta \boldsymbol{m} \cdot \boldsymbol{r}/r$, $\Delta m_B = \Delta \boldsymbol{m} \cdot \boldsymbol{B}/B$, $e_{rB} = \boldsymbol{r} \cdot \boldsymbol{B}/rB$. The spectral densities of correlation functions, $J_{\Delta m_\alpha \Delta m_\beta}$, can be expressed through the corresponding components of the dynamic susceptibility tensor (see Eq. (2)):

$$J_{\Delta m_\alpha \Delta m_\beta}(\omega) = \frac{2k_B T}{\omega} \operatorname{Im} \chi_{\alpha\beta}(\omega) \qquad (\hbar\omega << k_B T). \qquad (11)$$

The time evolution of the normalized muon polarization $P(t)=A(t)/A(0)$ is described by the expression

$$P(t) = \left\langle \frac{1}{N(x)} \sum_r \exp[-2w(\boldsymbol{r})t] \right\rangle_\mu \qquad (12)$$

where the sum is taken over $N(x)$ yttrium sites around the muon stopping site and $<...>_\mu$ means averaging over the different muon sites. Considering the average volume containing one holmium ion, we obtain a crude estimation of $N(x) \approx 1/x$.

In order to simulate the muon polarization $P$(t), in the present work we take an approach similar to that used to simulate spin-lattice relaxation measurements of the $^{19}$F nuclei in holmium doped LiYF$_4$ crystals.[8] We assume that in the case of a low concentration of Ho$^{3+}$ ions, each implanted muon interacts with only one Ho$^{3+}$ ion. We assume that the Ho$^{3+}$ ions are distributed uniformly in the host matrix; this approximation evidently fails for large $x > 0.01$. There are four possible muon stopping sites in the unit cell: $r_{1,2}=(a/2 \pm a/4\ -c/8)$ and $r_{3,4}=(\pm a/4\ a/2\ c/8)$.[17] So, the volume per one Ho$^{3+}$ ion contains $(2x)^{-1}$ unit cells and $N_m(x)=2/x$ possible muon stopping sites. We suppose that muons remain stationary without diffusion and neglect mixing of the muon spin states with spin states of neighboring F$^-$ nuclei in the magnetic fields $B > 0.01$ T.[17] Assuming independent relaxation of the muon polarization in different sites, we can describe the observed polarization decay by an expression similar to Eq. (12)

$$P(t) = \frac{1}{N_m(x)} \sum_r \exp[-2w(r)t] \qquad (13)$$

where the sum is taken over $N_m$ muon sites at the distances $r$ from the fixed Ho$^{3+}$ ion. In case of a large number of sites (small $x$) and $w(r) \propto 1/r^6$, a continuum approximation can be used that yields $P(x,B,T\,|\,t) = \exp\{-[\lambda(x,B,T)t]^{1/2}\}$.[25,26]

After the number of muon sites $N_m$ has been fixed, the muon polarization $P(t)$ is then calculated without introducing any additional parameters (except those which were used in simulations of the AC-susceptibilities). For a straightforward comparison of our simulations to the experimental data, the measured and calculated $P(t)$ curves (calculations were carried out for $0 \le t \le 50$ $\mu s$ using time increments of $\Delta t = 10^{-2}$ $\mu s$) were fit to a stretched exponential depolarization function $P(t) = \exp[-(\lambda t)^\beta]$. The exponent $\beta$ was found to vary weakly in the range $0.5 \le \beta \le 0.8$. These values are consistent with a depolarization dominated by fluctuating dilute magnetic ions.[25,26] An example of the measured depolarization data, along with the corresponding curve fits and simulated curves, is shown in Fig. 1. The $\chi^2$ per degree of freedom was between 1.02 and 1.4. However, it should be noted that we obtain an overestimate of the decrease of the polarization at short times using the stretched exponential functions. To analyze dependences of the relaxation rate $\lambda$ on temperature and magnetic field, as a rule, we used stretched exponential functions with the fixed exponent $\beta = 1/2$ to fit the simulated curves. The final values of parameters $N_m(x)$ used in simulations were obtained from fitting the maximum value of $\lambda(x,B,T)$ vs temperature in the magnetic field $B = 0.023$ T parallel to the $c$-axis.

## IV. AC-SUSCEPTIBILITY RESULTS
### A. Frequency Dependent Susceptibility

We measured the frequency dependence of the dynamic susceptibility at a fixed field of 38.5 mT (to avoid features at whole or half integer $n$ values) and $T$ = 1.75 K for $x$ = 0.0017 and 0.0085 (shown in Fig. 2). For $x$ = 0.0017, $\chi'$ and $\chi''$ agree with the results of Bertaina et al.[6] at all measured frequencies. A maximum in $\chi''$ and roll-off in $\chi'$ centered at approximately 400 Hz identify the characteristic frequency $f_0$ of fluctuations of the holmium magnetic moment. Data for the $x$ = 0.0085 sample agree with the data for $x$ = 0.0017, with magnitude (in units of emu/g) differing by the concentration scaling factor 5, for $\chi'$ up to approximately 400 Hz, and then deviates qualitatively from the results for $x$ = 0.0017 at higher frequencies. There is a significant roll-off in $\chi'$ that has not saturated by 10 kHz; similarly $\chi''$ increases monotonically with no maximum reached by 10 kHz. While these features dominate the behavior of the curve, we note that a small inflection (shoulder) in $\chi'$ ($\chi''$) appearing at ~ 400 Hz (similar to the features observed for $x$ = 0.0017 at the same frequency), suggesting more than one characteristic frequency of spin fluctuations at this concentration.

The in-phase and out-of-phase susceptibilities vs. frequency were calculated according to Eq. (2). We used previously given parameters of the electron-deformation interaction[6] when calculating relaxation rates of the holmium magnetization but varied phonon bottleneck factors $P_{mn}$ and parameters of the cross-relaxation (widths $\Delta$ of the form-function and factors $\delta$ and $\varepsilon$, see Eqs. (6) and (7)). The final values of the parameters which were cross-checked by the analysis of the magnetic field dependences of the susceptibility (see below) are presented in Table I. Changes of the parameters with the holmium concentration are discussed in the next section. The simulated frequency dependences for $x$ = 0.0017 match well the experimental data (see Fig. 2a) in the region of low frequencies ($\omega < f_0$). However, at higher frequencies, the measured in-phase susceptibility exceeds the calculated values, but we cannot exclude that this is an experimental artifact. For $x$ = 0.0017, the position of the maximum of $\chi''$ is determined by the interaction of the $Ho^{3+}$ ions with the phonon bath, but in the sample with $x$ = 0.0085 the calculated frequency of the main maximum ($f_0$ = 14 kHz) is determined primarily by the cross-relaxation rates which become effective not only at ALCs but in between ALCs as well due to the large width of the cross-relaxation form-function.

## B. Field Dependent Susceptibility: x = 0.0017 and x = 0.0085

The measured magnetic field dependences of $\chi'$ and $\chi''$ at constant frequencies and temperature $T = 1.9$ K for $x = 0.0017$ and $T = 1.8$ K for $x = 0.0085$ are shown in Figs. 3 and 4, respectively. At field values corresponding to ALCs, cross-relaxation processes are effective while in between the crossings the cross-relaxation processes are negligible provided the dispersion $\Delta$ is small. Measurements for $x = 0.0017$ showed characteristic peaks (dips) in $\chi'$ ($\chi''$) at resonant field values $B_n$ ($-7 \leq n \leq 7$), an additional peak at $n = 8$, and smaller features at half-integer $n$ values, all agreeing with the previous work of Refs. 6,7,13. As we increased the measurement frequency, we observed a suppression of the small peaks at half-integer $n$ in $\chi'$, and a 'dip-to-peak' transition in $\chi''$ between 800 and 3000 Hz, again agreeing with observations of Bertaina et al.[6] indicating that the measurement frequency has increased above the effective relaxation rate.

The field dependence of $\chi'$ and $\chi''$ in the sample with the 5-times increased concentration $x = 0.0085$ (see Fig. 4) exhibits subtle but important differences compared to the data for $x = 0.0017$. Peaks (dips) in $\chi'$ ($\chi''$) continue to occur at ALCs for $-7 \leq n \leq 7$ and $n = 8$, however at 800 Hz the features are significantly broader than those seen in the $x = 0.0017$ sample. These become sharper as frequency is increased, and at 10 kHz we see a flattening of the regions between the dips in $\chi''$. At 100 kHz the amplitude of the sharp peaks at $B_n$ values is larger for $n = 1, 3, 5$ which correspond to the three largest tunnel splittings at the ALCs, as predicted by Giraud et al.[5] and seen by Graf et al.[20] in transverse field μSR measurements. The half-integer peaks in the field dependences induced by co-tunneling processes are no longer observed, and this is reproduced in the simulations through increased broadening of the cross-relaxation form-function.

There are two types of 'dip-to-peak' crossover observed in $\chi''$ at concentration $x = 0.0085$. One, also seen in the $x = 0.0017$ sample, occurs at all ALC field values $B_n$. For $x = 0.0085$ these occur at a frequency of 18 kHz, which is 6-times the crossover frequency found for $x = 0.0017$, close to the concentration ratio of 5, and consistent with the prediction that relaxation rates increase with increasing concentration. The other is a crossover of a much broader feature superimposed on the ALC signatures, and centered at zero field. At 800 Hz, $\chi''$ has a dip at zero field with peaks located at ± 0.2 T, and with increasing frequency these develop into a broad peak at zero field by 10 kHz. The crossover seems to occur near 3 kHz. To reproduce these broad features in our simulations, we had to increase the width $\Delta$ of the cross-relaxation form-function linearly with $x$ and, according to trends found earlier, to compensate this broadening by also increasing the cross-relaxation strength parameter $\delta$, indicating that cross-relaxation processes are no longer negligible at field values in between the resonant fields $B_n$. Such a behavior of the cross-relaxation parameters agrees with the frequency dependences of $\chi'$ and $\chi''$ described in the preceding section and with the physical nature of the cross-relaxation processes. It is also interesting that we had to *decrease* the phonon relaxation rates by increasing the phonon bottleneck factors $P_e$ and $P_g$ in the sample with $x = 0.0085$ (see Table I). In Refs. 6 and 7, $P_e$ and $P_g$ were found to anomalously decrease with increasing $x$ (for very low $x$ values, $x \leq 0.003$); the return of the expected increase in the $P_{mn}$ factors with $x$ may be caused by non-linear complex dependences of the transition widths $\Delta\omega_{mn}$ and phonon lifetimes $\tau_{ph}(\omega_{mn})$ on the concentration $x$ of impurity holmium ions. The parameters presented in Table I may have relatively small uncertainties caused by the misalignment of the samples, since the effective relaxation rate increases as the magnetic field declines from the $c$-axis. In particular, according to calculations, the position of the maximum of $\chi''(\omega)$ in Fig. 2a shifts to higher frequencies by about 40 Hz in the magnetic field declined by 6 degrees from the $c$-axis, and also the peaks at ALCs in the field dependences of $\chi'$ become slightly broader. It should be noted that these effects are practically independent of the direction of the field projection on the $ab$-plane.

### C. Field Dependent Susceptibility: $x$ = 0.0408 and 0.0855

Susceptibility measurements were also made at fixed temperature (1.8 K) and frequencies for the $x$ = 0.0408 and 0.0855 concentrations, shown in Fig. 5. We observe broad structures that evolve with frequency at high fields (relative to $x$ = 0.0017 and 0.0085) and may indicate additional relaxation processes that are effective in clusters of the paramagnetic ions with different geometries.[27] However, for $x \leq 0.1$, the probability of formation of dimer centers involving two closely spaced holmium ions (such a pair should exhibit essential changes of its spectral and kinetic properties relative to the properties of a single ion) is very small,[28] and we assume that the utilization of the single ion model with renormalized parameters is still appropriate, at least, at high enough temperatures so that quantum correlations are destroyed by thermal excitations.

In the $x = 0.0408$ sample the in-phase susceptibility $\chi'$ begins at low frequency as a broad peak with a shoulder at ~ 0.25 T, and evolves into a single peak at 100 kHz. The out-of-phase susceptibility $\chi''$ shows what looks to be two shoulders at 0.2 and 0.4 T that combine into one at 100 kHz. A pronounced dip in $\chi''$ in zero field at 100 Hz crosses over at about 3 kHz and becomes a peak at 100 kHz. Similar features appear present in the $x = 0.0855$ sample. $\chi'$ begins as a wide peak and sharpens with increased frequency, with a well defined shoulder that persists up to 10 kHz. $\chi''$ has three 'features' that converge with increased frequency to a peak at ~ 0.3 T and 10 kHz. A dip-to-peak crossover also occurs at zero field for this concentration at 3 kHz. Finally, we also note that the $\chi''$ data for $x = 0.0408$ and $0.0855$ appear to have similar field dependences in the high frequency limit, with peaks at zero field and at ± 0.2 T and ± 0.3 T, respectively.

The results of simulations reproduce satisfactorily the low frequency susceptibilities of the sample with $x = 0.0408$ (see Figs. 5a and 5b), in particular, the low-field peaks in $\chi''$ and the broad structure dip-to-peak crossover (attributed, as in the case of $x = 0.0085$, to Ho-Ho cross relaxation) centered at zero field which occurs at ~ 3 kHz. However, though single-ion physics modified to include spin-spin relaxation remains important even in the higher concentrations, a comparison of the theoretical and experimental data shows that large features of the measured dynamic susceptibilities in the samples with $x = 0.0408$ and $0.0855$ (in particular, the additional high-field peaks of $\chi''$) are presumably caused by collective behavior of paramagnetic ions.

Frequency and temperature dependences of the dynamic susceptibility of LiHo$_x$Y$_{1-x}$F$_4$ samples with the holmium concentration $x = 0.045$ were studied at low temperatures (50 mK ≤ $T$ ≤ 350 mK) and in zero DC magnetic field in Refs. 3, 10, 29, and 30, and more recently in Ref. 11 where samples with $x = 0.018$ and $x = 0.08$ were studied as well. Calculations of the frequency dependence of $\chi'$ and $\chi''$ in this range of temperatures using the parameters determined from measurements at higher temperatures (Table I) give evidence for underestimated relaxation rates of the Ho$^{3+}$ ions. In particular, the frequencies associated with the position of the maximum of $\chi''(\omega)$, $f_0$, obtained from calculations for the sample with $x = 0.0408$ at temperatures 0.2 - 0.25 K were found to be in the range 20 to 40 Hz, smaller by an order of magnitude than the corresponding measured values for the sample with $x = 0.045$, and even reduced by a factor of two from the measured $f_0$ for the sample with $x = 0.018$. Because the phonon relaxation is suppressed at sub-Kelvin temperatures, we have only one free parameter, the cross-relaxation line width $\Delta$, that can be varied to achieve an agreement between the experimental and calculated relaxation rates. Our simulations, for which we use the parameters from Table I for $x = 0.0408$ except that the line width $\Delta$ is reduced to 400 MHz due to the expected narrowing at low temperatures, reproduce accurately the frequency dependences of the real and imaginary parts of the dynamic susceptibility presented in Refs. 10 and 11 for the sample with $x = 0.045$ at $T = 0.2$ K. However, with decreasing temperature, the calculated frequency $f_0$ of the magnetization fluctuations follows an Arrhenius law with small activation energy of about 0.6 K (comparable with the hyperfine splitting of the ground doublet of the holmium ions) and exceeds the observed $f_0$ by two orders of magnitude at $T = 80$ mK. A critical slowing down of spin fluctuations is expected upon approaching the estimated spin freezing temperature of about 60 mK for $x = 0.045$, and this effect is responsible for the difference between the measured and simulated data.

## V. MUON SPIN RELAXATION RESULTS
### A. Temperature dependence of relaxation rates

Analysis of the muon relaxation rates $\lambda$ in the sample with the lowest concentration $x = 0.0017$ shows that we are able to reproduce, at least qualitatively, the observed dependences of $\lambda$ on temperature and magnetic field in the framework of the model described in Section III.B using parameters determined from the susceptibility simulations and only one additional parameter, the number of muon sites $N_m$ per one impurity paramagnetic ion, which assumes a physically reasonable value. Figure 6 presents calculated values of $\lambda$ for $N_m = 780$ ($r \leq 1.892$ nm) that match well the results of measurements in the magnetic field of 23 mT at the ALC. At higher fields, the calculated relaxation rates decrease, in agreement with the experimental data, although, the theoretical rates are overestimated by about 50% for a magnetic field of 60 mT (located in between the ALCs).

In Fig. 7 we compare measured temperature dependent relaxation rates $\lambda$ for $x = 0.0085$ with those extracted from simulations, with the field (and therefore muon polarization oriented both parallel and perpendicular to the $c$-axis. Simulations were carried out taking using $N_m = 248$ muon sites ($r \leq 1.268$ nm) in the sum Eq. (13). There is remarkable agreement between the measured and calculated rates for the magnetic field $B = 23$ mT parallel to the $c$-axis (at the first ALC) and for magnetic fields 23 and 60 mT perpendicular to the $c$-axis. It should be noted that the calculated relaxation rates increase non-monotonically in the magnetic field rotating from the $c$-axis to the $ab$-plane, the dependence of $\lambda$ on the angle $\vartheta$ between the field and the $c$-axis exhibits a narrow dip with the width of about eight degrees where $\lambda(\vartheta = 90°)$ is less by about 20% than the maximum values at $\vartheta = 90° \pm 4°$. The relaxation rates for $x = 0.0085$ are about an order of magnitude larger than for $x = 0.0017$ at the same field and temperature, and the temperature dependences of $\lambda$ in both samples are qualitatively similar, with a maximum near 10 K, which is close to the value of the energy barrier to spin reversal between the ground state doublet and the first excited singlet. This thermally activated relaxation is similar to what was observed in the spin-lattice relaxation rate of fluorine nuclei via NMR[31] for $x = 0.00127$ and successfully simulated in Ref. 8; the maximum in the temperature-dependent rate is characteristic of a slowing down of thermally driven magnetic fluctuations on the muon timescale.

Figure 8 shows muon relaxation rates for $x = 0.0408$ and 0.0855. In contrast to the data at lower concentrations, the low field rate monotonically increases with decreasing temperature, approaching a constant value as $T \to 0$ K. The saturation is consistent with data reported in Ref. 32 on samples with comparable concentrations, which were interpreted as resulting from persistent low temperature fluctuations of local magnetic fields described by a single correlation time. We find that strong external fields begin to suppress the low temperature relaxation rate, with stronger fields required for the sample with higher Ho concentration. We calculated the muon relaxation rates using the number of muon sites $N_m = 52$ ($r \leq 0.73$ nm) in Eq. (13) and the parameters of the cross-relaxation which were found from the analysis of the low frequency susceptibility for $x = 0.0408$ (see Table I). The results agree well with the measured rates at temperatures $T > 6$ K, when the fluctuations of single holmium magnetic moments play a dominant role. However, at low temperatures the calculated rates of the muon polarization decay are significantly underestimated, indicating the failure of our model in this regime. The precession of the muon magnetic moments in quasi-static random magnetic fields may be a source of the observed fast decay of the initial muon polarization at temperatures below 6 K in the samples with holmium concentrations $x \geq 0.04$, since our single-ion model clearly cannot account for disorder. Stochastic theory[33] shows that an external field suppresses the polarization decay due to random fields that is consistent with our findings. However, the fields required to begin suppressing $\lambda$ (roughly 600 G or more) at these concentrations are significantly larger than $\lambda/\gamma_\mu$ (of order 50 G), so we conclude that dynamical fluctuations must play an important role at low temperatures. Moreover, since our model accounts for single-ion fluctuations, these must be related to collective behavior of the $Ho^{3+}$ ions, possibly due to incomplete freezing of the spin glass state, as proposed in Refs. 9 and 32 and described theoretically in Ref. 12.

### B. Field dependence of relaxation rates

Closer inspection of the data in Fig. 8 reveals an unusual behavior: in the temperature interval 10 K < $T$ < 20 K the muon relaxation rate in strong fields is *higher* than the observed values in lower fields (for ***B*** || *c* only). To probe this effect further we conducted more detailed field dependent measurements in this temperature interval on samples with $x$ = 0.0085 and 0.0855. Data were taken at both ISIS (to 0.25 T) and PSI (to 0.6 T) at fixed temperature $T$ = 12 K, confirming the increasing relaxation rate with increasing field. To follow the behavior to higher fields, we performed measurements on the HiFi spectrometer at ISIS up to 3 T on $x$ = 0.0085 ($T$ = 16 K) and 0.0855 ($T$ = 12 K, 16 K, and 20 K) and observed a peak between 0.5 and 1 T that shifts to higher fields with increasing temperature, as shown in Fig. 9(a) and its corresponding inset. All the data collapse onto a single curve by scaling the relaxation rate and magnetic field to the peak values $\lambda_0$ and $B_0$, respectively, as shown in Fig. 9(c), indicating that the origin of the behavior lies in single-ion physics rather than any unusual collective behavior.[34] While the value of $\lambda_0$ at 16 K increases by a factor of twenty (from 0.063 μs$^{-1}$ to 1.18 μs$^{-1}$) upon increasing $x$ by a factor of ten, $B_0$ increases weakly, by a factor of two (from 3000 G to 6000 G).

A possible explanation for the unusual field dependent depolarization rate may be related to an anomaly in the phonon density of states (PDS). The calculated PDS of LiYF$_4$ has a relatively broad maximum[35] at phonon energies 50-70 cm$^{-1}$ (see Inset in Fig. 9(b)). The position of this maximum is very close to the crystal field level $\Gamma_{3,4}^2$ of the $Ho^{3+}$ ions in LiYF$_4$ (72 cm$^{-1}$).[36] Due to the specific shape of the PDS, the contribution of Orbach processes to the relaxation rate of the $Ho^{3+}$ magnetic moment at temperatures $T$ > 10 K depends strongly on the differences between the energies of the hyperfine sublevels of the excited $\Gamma_{3,4}^2$ and ground $\Gamma_{3,4}^1$ doublets. For a qualitative picture, we assume a model of dynamical depolarization and approximate the transition probability (8) by the expression

$$w \propto |V|^2 \frac{\Gamma}{\omega_\mu^2 + \Gamma^2} \qquad (14)$$

where $hV$ is average energy of interaction between a muon and holmium ions. For a $Ho^{3+}$ fluctuation rate $\Gamma$ larger than the muon Larmor frequency $\omega_\mu$, increasing the magnetic field parallel to the *c*-axis will cause transition energies between the electron-nuclear states to shift out of the region where the PDS has the maximum, the fluctuation rate of magnetic moment decreases, and the muon depolarization rate can then increase. Note that we are considering here temperatures above the location of the maximum in the temperature dependence of the muon relaxation rate. We have simulated the field dependence of the depolarization rate in the sample with $x = 0.0085$ (see Fig. 9(b)) by making use of the PDS (shown in the inset) calculated in the framework of the rigid ion model of lattice dynamics of $LiYF_4$; it should be noted that spectral densities of autocorrelation functions for differences between displacements of the $Y^{3+}$ ion and the nearest neighbor $F^-$ ions also contain well pronounced maxima at the frequencies from 60 to 70 cm$^{-1}$ (Fig. 3 in Ref. 37). The general behavior – an initial decrease in $\lambda$ with field, followed by a peak at about 0.3 T, for ***B*** along the *c*-axis, but a monotonically decreasing rate with *B* perpendicular to it – is indeed reproduced by the simulations, although the magnitude of the measured effect is much larger than that predicted by the simulations. Thus we believe that this model captures the underlying physical process, but a more refined calculation of coupling constants in the Hamiltonian of electron-phonon interaction is required to obtain detailed quantitative agreement.

## VI. DISCUSSION AND SUMMARY

We have successfully simulated the results of magnetic susceptibility and µSR measurements performed on samples with $Ho^{3+}$ concentrations up to $x = 0.0085$, confirming that at these low concentrations the relaxation rates increase with *x* in the temperature range 1.75 – 4 K (seen in previous work[6] as shifts to higher frequencies of the maxima in frequency dependences of $\chi''$ ($\omega$)) and find that contributions to the relaxation shift from being dominated by the electron-phonon interaction to cross relaxation as correlations between the $Ho^{3+}$ ions become important. We note that by definition the bottleneck factors should increase linearly with concentration *x*, and this is in agreement with our measurements. The observed earlier[6-8] reduction of these factors with $x$ for $x < 0.003$ can be the result of the nonlinear increase of transition widths $\Delta\omega_{mn}$ with concentration in the range of small $x$. Cross relaxation between pairs of the Ho$^{3+}$ ions has been taken into account in our single-ion picture, but could hypothetically be extended to $N$-ion co-tunneling processes with $N > 2$. This suggests a different type of collective mechanism, such as fast relaxation processes between clusters of ions like those mentioned by Stevens.[27] The corresponding calculations are necessary to make clear the origin of features seen at high fields and high frequencies in susceptibility measurements at concentrations $x \geq 0.0408$.

Calculated rates of muon depolarization agree qualitatively with the data at different temperatures and in different magnetic fields parallel and perpendicular to the crystal symmetry $c$-axis. Still, differences between the calculated and measured magnitudes of $\lambda$ are not negligible. Parameters of the model were determined from fitting the AC-susceptibility measured in the magnetic fields parallel to the crystal symmetry axis and at frequencies not exceeding $10^5$ Hz. However, the measured muon spin relaxation rates are connected with the responses of the Ho$^{3+}$ ions at frequencies up to 100 MHz and not only in the fields parallel to the $c$-axis, but perpendicular to this axis as well. It should be noted that, according to the results of our calculations, the relaxation rate $\lambda$ depends remarkably on the orientation of the magnetic field. In particular, at fixed temperature, $\lambda$ increases when the field rotates from the $c$-axis to the basal plane and has maximal values in fields tilted from this plane by $\pm\ 4^o$. For the magnetic field $\boldsymbol{B}$ exactly normal to the $c$-axis, the rate $\lambda$, considered as a function of the angle $\vartheta$ between the field and the $c$-axis, has a local minimum at $\vartheta = 90^o$, although the calculated relative differences between the maximal and minimal values, $1 - \lambda(\vartheta = 90^o)/\lambda(\vartheta = 90 \pm 4^o)$, do not exceed 20%. This dependence confirms the important role of fluctuations of the holmium magnetic moment, which are enhanced due to the electron-nuclear ALC induced by a transverse field.

The derived method of calculation of the muon spin relaxation in strong longitudinal fields, although shown here to be adequate up to $x = 0.0085$, is based on several crude assumptions: a model of single-ion magnets was employed, with a homogeneous distribution of impurity paramagnetic ions supposed; also, possible variations of phonon bottleneck factors with temperature, Raman relaxation processes, and local lattice deformations caused by implanted muons were all neglected. Nevertheless, it may serve as a basis for more elaborate theoretical studies of the muon relaxation in samples with higher concentration of paramagnetic ions.

## ACKNOWLEDGMENTS

This work was supported by NSF grant DMR-0710525. Experiments were performed at the ISIS Muon Facility at the Rutherford Appleton Laboratories (UK) and the Swiss Muon Source at the Paul Scherrer Institute (Switzerland). We gratefully acknowledge the contribution of Dr. Alexandra Tkachuk who provided the samples studied in this work.

* corresponding author (grafm@bc.edu)

TABLE CAPTIONS

Table I. Parameters of the phonon bottleneck and cross relaxation in $LiY_{1-x}Ho_xF_4$ crystals. The results of the present work are in bold, while other values are taken from Refs. 6,7, and 8, where they have been determined from studies of the AC-susceptibilities and fluorine nuclear relaxation rates. $\alpha = -1/450$ is the reduced matrix element for 2nd rank spherical tensor operators.

TABLES

Table I

|  | Units | Concentration of $Ho^{3+}$ ions (in %) | | | | | | |
|---|---|---|---|---|---|---|---|---|
|  |  | 0.104 | 0.13 | 0.157 | **0.17** | 0.27 | **0.85** | **4.08** |
| $P_e$ | $(2\pi \cdot 10^9)^2$ s$^{-1}$ | 5.2 | 4.4 | 3.6 | **2.5** | 2.0 | **4** | **12** |
| $P_g$ | $(2\pi \cdot 10^9)^2$ s$^{-1}$ | 55 | 45 | 36 | **26** | 18 | **50** | **120** |
| $\Delta$ | MHz | 185 | 185 | 220 | **400** | 240 | **2000** | **2200** |
| $\delta/h$ | $10^7$ s$^{-1}$ | 7.0 | 7.73 | 8.6 | **4.85** | 11.2 | **13.7** | **34.3** |
| $\varepsilon$ | $\alpha^{-2}$ | 0.04 | 0.04 | 0.04 | **0.013** | 0.04 | **0.025** | **0.025** |

FIGURE CAPTIONS

FIG. 1. (a) Example curves showing time dependent asymmetry of muon decays measured at ISIS for two sample concentrations $x = 0.0085$ and $x = 0.0855$ in longitudinal field $B = 23$ mT parallel to the $c$-axis, at $T = 8$ K. Solid lines are fits to stretched exponential functions $23.3\exp[-(0.2\ t/\mu s)^{0.5}]$ and $19\exp[-(2.8\ t/\mu s)^{0.65}]$, respectively. (b) Example simulated curves for time dependent muon polarization in longitudinal field $B = 60$ mT parallel to the $c$-axis at 10 K for $x = 0.0017$ and $x = 0.0085$ as compared to the measured data presented by symbols. Both curves can be approximated by the stretched exponential functions with the same value of $\beta = 0.55$ and $\lambda = 0.0075\ \mu s^{-1}$, $\lambda = 0.155\ \mu s^{-1}$, respectively.

FIG. 2. Measured frequency dependences of $\chi'$ (solid symbols) and $\chi''$ (open symbols) compared with calculation (solid line) at 1.75 K in a constant 38.5 mT field for (a) $x = 0.0017$ and (b) $x = 0.0085$.

FIG. 3. Measured (bold lines) and calculated (thin lines) susceptibilities $\chi'(\omega)$ and $\chi''(\omega)$ of LiY$_{1-x}$Ho$_x$F$_4$ ($x = 0.0017$) at $T = 1.9$ K and constant frequencies; 0.8 kHz, 3 kHz (+8, +3), 5 kHz (+16, +6.5), 10 kHz (+23, +10.5). Curves are offset for clarity, and the numbers in parentheses represent shifts of the corresponding curves in units of emu/mol for $\chi'$ and $\chi''$, respectively.

FIG. 4. Measured (bold lines) and calculated (thin lines) susceptibilities $\chi'(\omega)$ and $\chi''(\omega)$ of LiY$_{1-x}$Ho$_x$F$_4$ ($x = 0.0085$) at $T = 1.8$ K and constant frequencies; 0.8 kHz, 3 kHz (+4, +3.5), 10 kHz (+8, +4), 18 kHz (+16, +8), 32 kHz (+24, +11), 56 kHz (+31, +14). Curves are offset for clarity, see Fig. 3 caption.

FIG. 5. Measured (bold lines) and calculated (thin lines) susceptibilities $\chi'(\omega)$ and $\chi''(\omega)$ of LiY$_{1-x}$Ho$_x$F$_4$ for $x = 0.0408$ (a, b) and $x = 0.0855$ (c, d) at $T = 1.8$ K and constant frequencies 0.1 kHz, 0.8 kHz (+3.5, +2), 3 kHz (+6, +3), 18 kHz (+8.5, +4), 32 kHz (+10, +4.5), 56 kHz (+11.5, +5), 100 kHz (+13, +5.5) (a, b) and 0.1 kHz, 0.8 kHz (+1.5, +0.2), 3 kHz (+3, +0.5), 5 kHz (+4.5, +0.7), 7 kHz (+6, +0.8), 10 kHz (+7.5, +1.0) (c, d). Curves are offset for clarity, see Fig. 3 caption.

FIG. 6. Measured (symbols) and calculated (solid lines) temperature dependences of the muon relaxation rate in the sample LiHo$_x$Y$_{1-x}$F$_4$ ($x = 0.0017$) at two magnetic fields, 23 mT (1, circles) and 60 mT (2, squares), parallel to the $c$-axis.

FIG. 7. Measured (symbols) and calculated (solid lines) temperature dependences of the muon relaxation rate in the sample LiHo$_x$Y$_{1-x}$F$_4$ ($x = 0.0085$) at two magnetic fields, 23 mT (circles) and 60 mT (squares), parallel (a) and perpendicular (b) to the $c$-axis.

FIG. 8. Measured (symbols) and calculated (solid lines) temperature dependences of the longitudinal relaxation rate in the samples LiHo$_x$Y$_{1-x}$F$_4$ ($x = 0.0408$ (a) and 0.0855 (b)) at fields 23 mT (circles), 60 mT (squares), and 120 mT (triangles) parallel to the $c$-axis.

FIG. 9. (a) Measured field dependence of the longitudinal relaxation rate $\lambda$ for $x = 0.0855$ at $T = 12, 16$, and 20 K. (Inset a) Measured field dependence of $\lambda$ for $x = 0.0085$ at $T = 16$ K. (b) Simulated field dependence of $\lambda$ for $x = 0.0085$ at $T = 16$ K with the $c$-axis parallel (solid curve) and perpendicular (dashed curve) to $B$. (Inset b) Calculated phonon density of states $\rho(\omega)$ for LiYF$_4$. (c) This plot shows scalability of $\lambda$ for both concentrations $x = 0.0085$ and $x = 0.0855$ at all temperatures, data for $x = 0.0085$ at 16 K are shown as open diamonds.

FIGURES

Figure 1

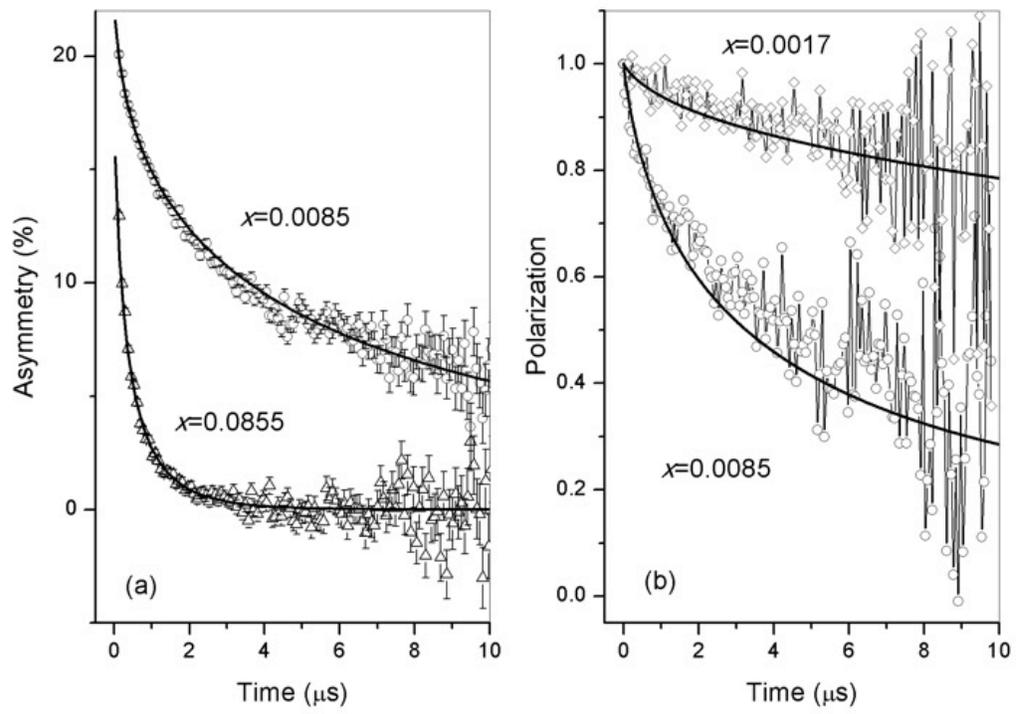

Figure 2

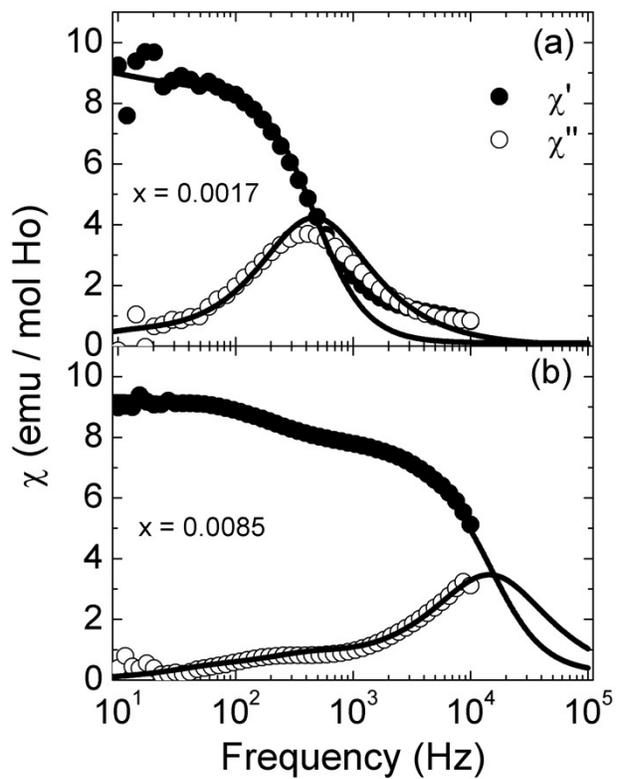

Figure 3

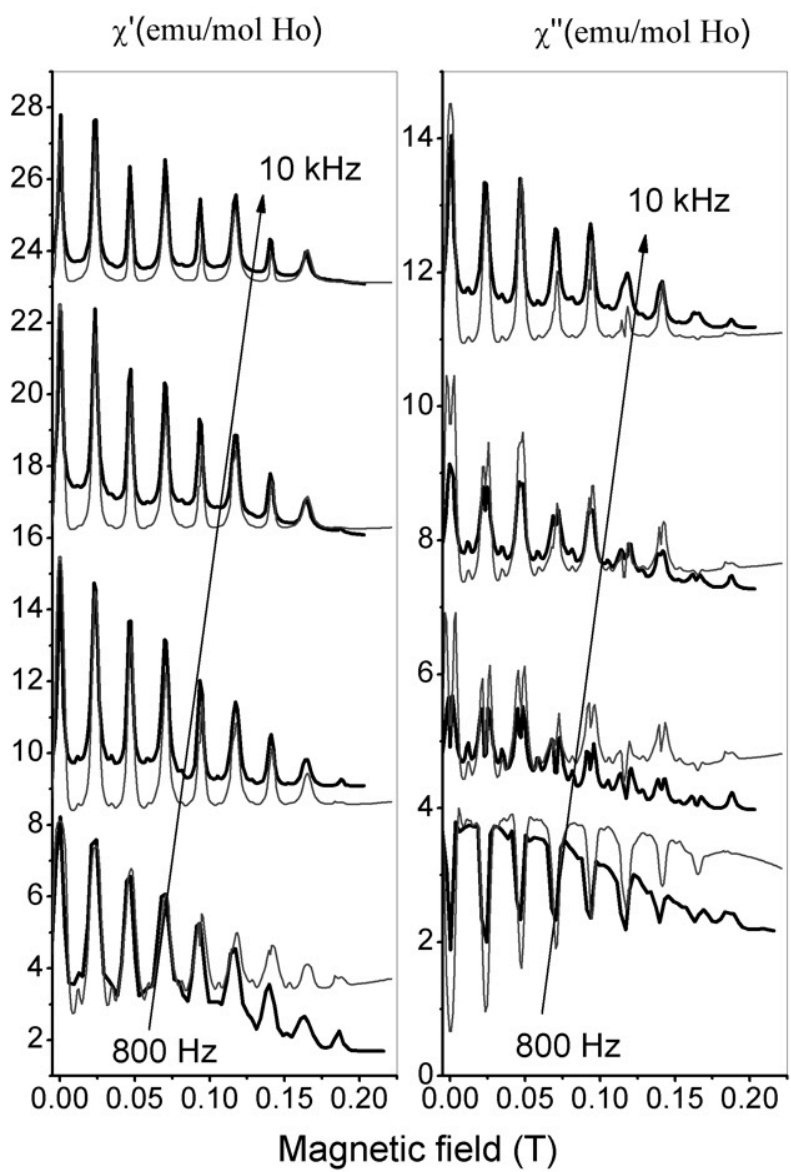

Figure 4

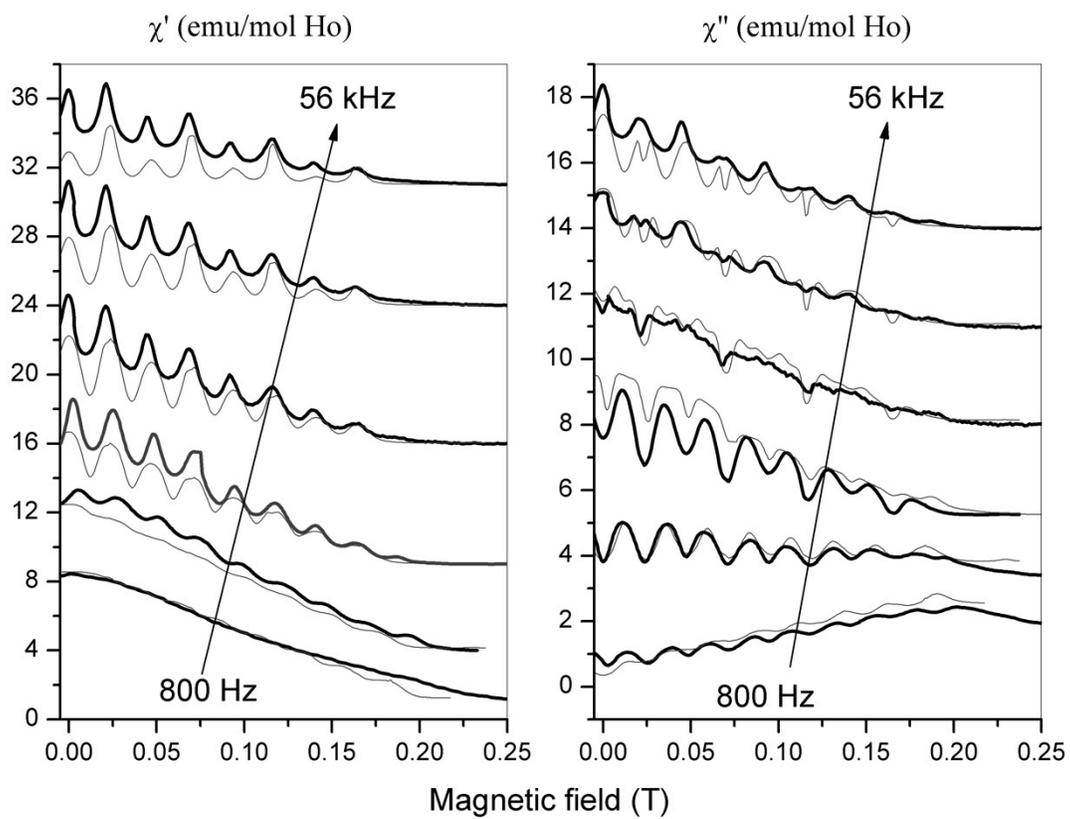

Figure 5

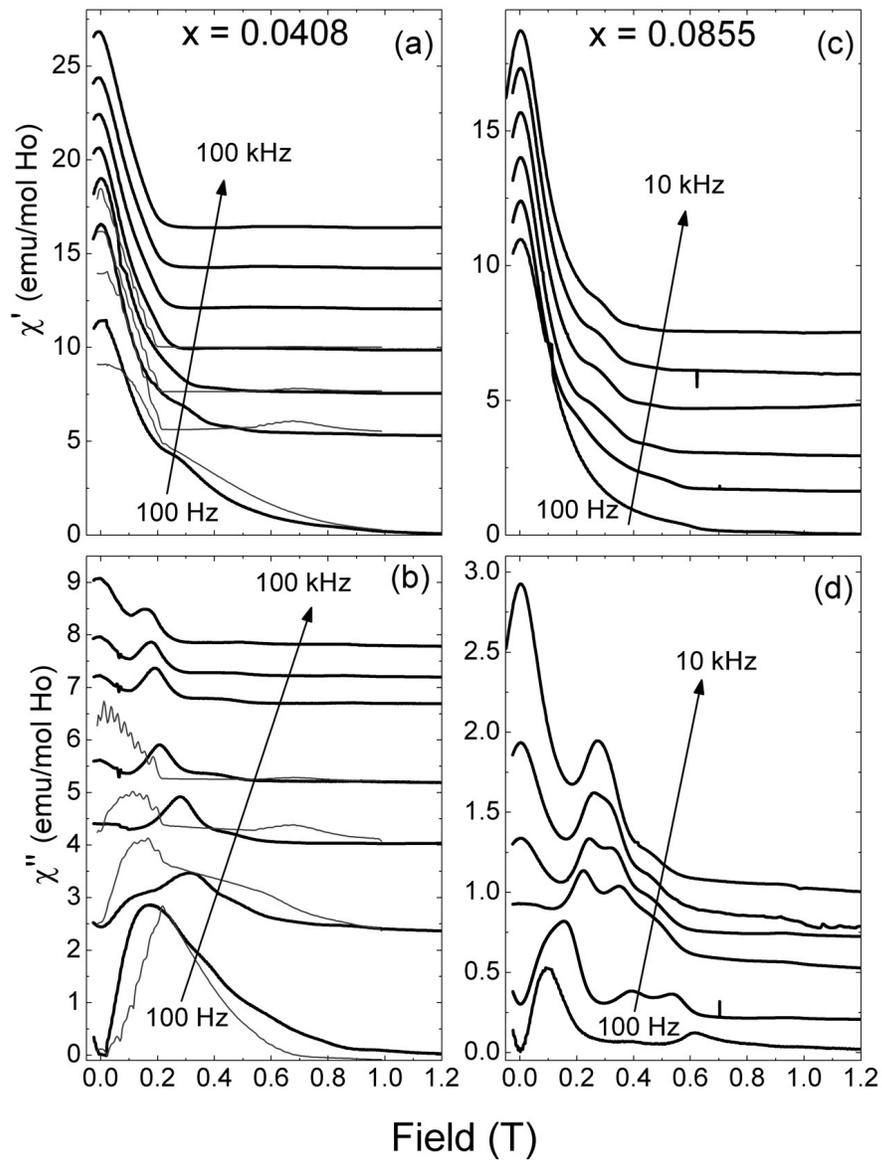

Figure 6

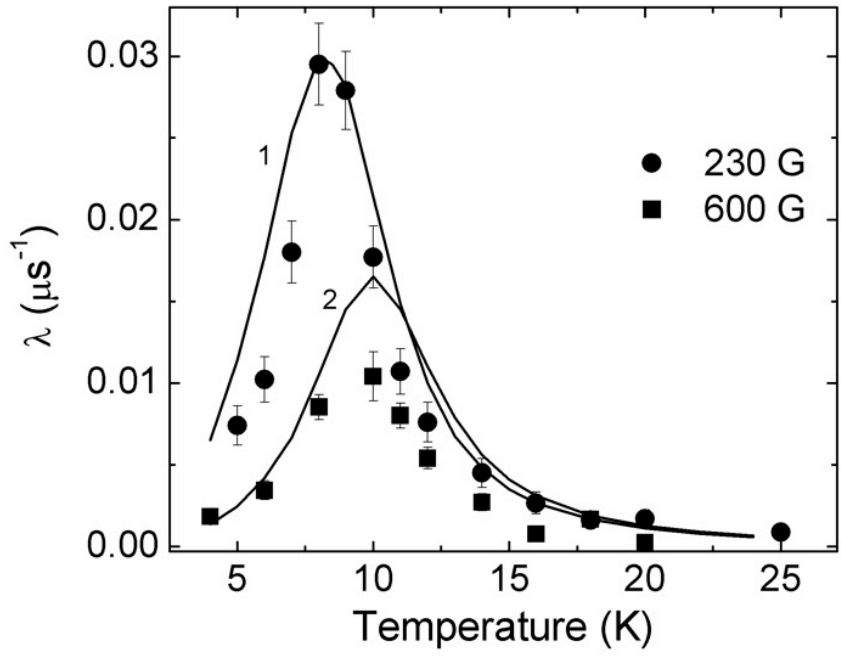

Figure 7

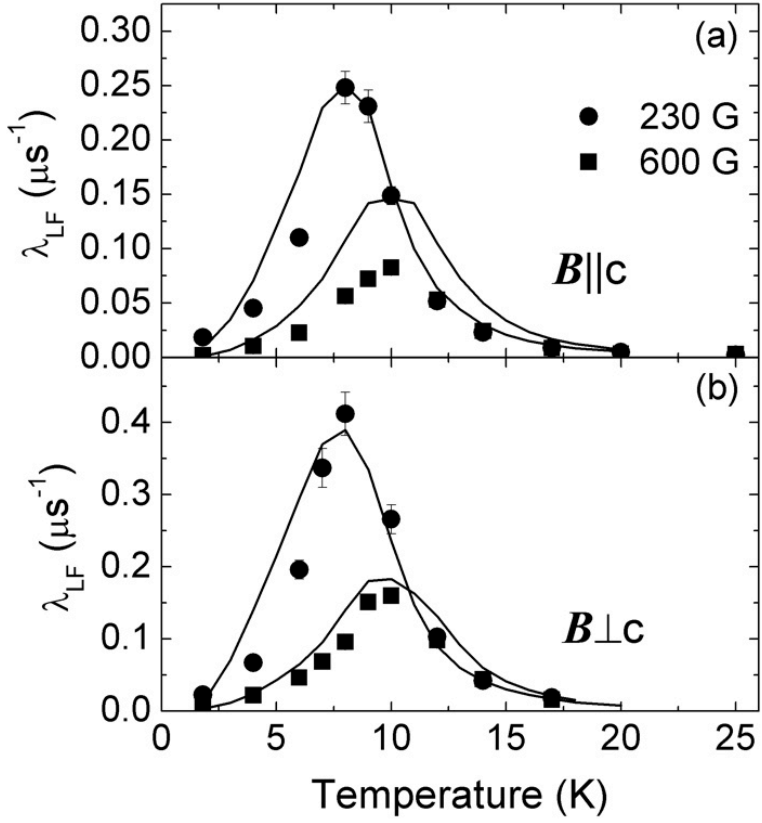

Figure 8

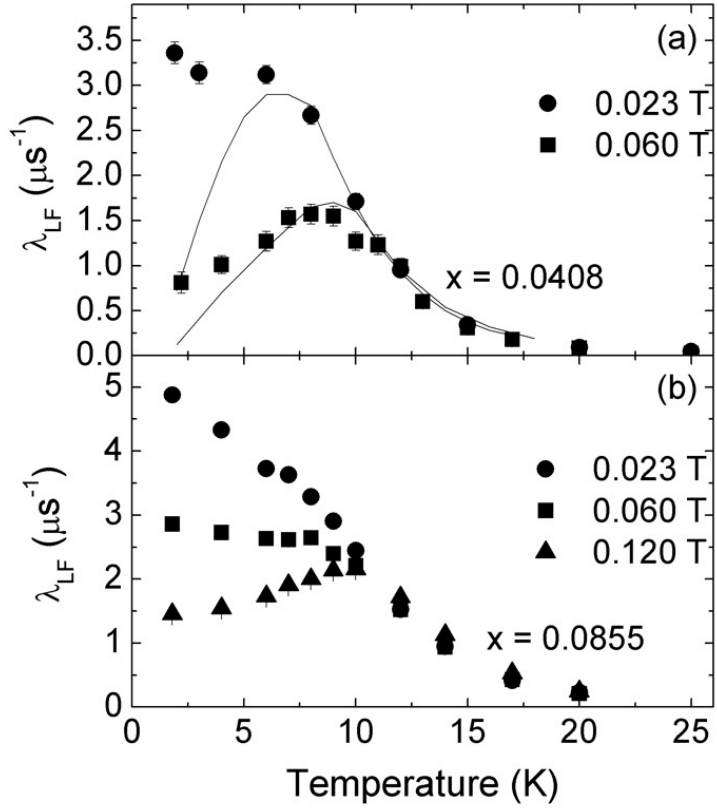

Figure 9

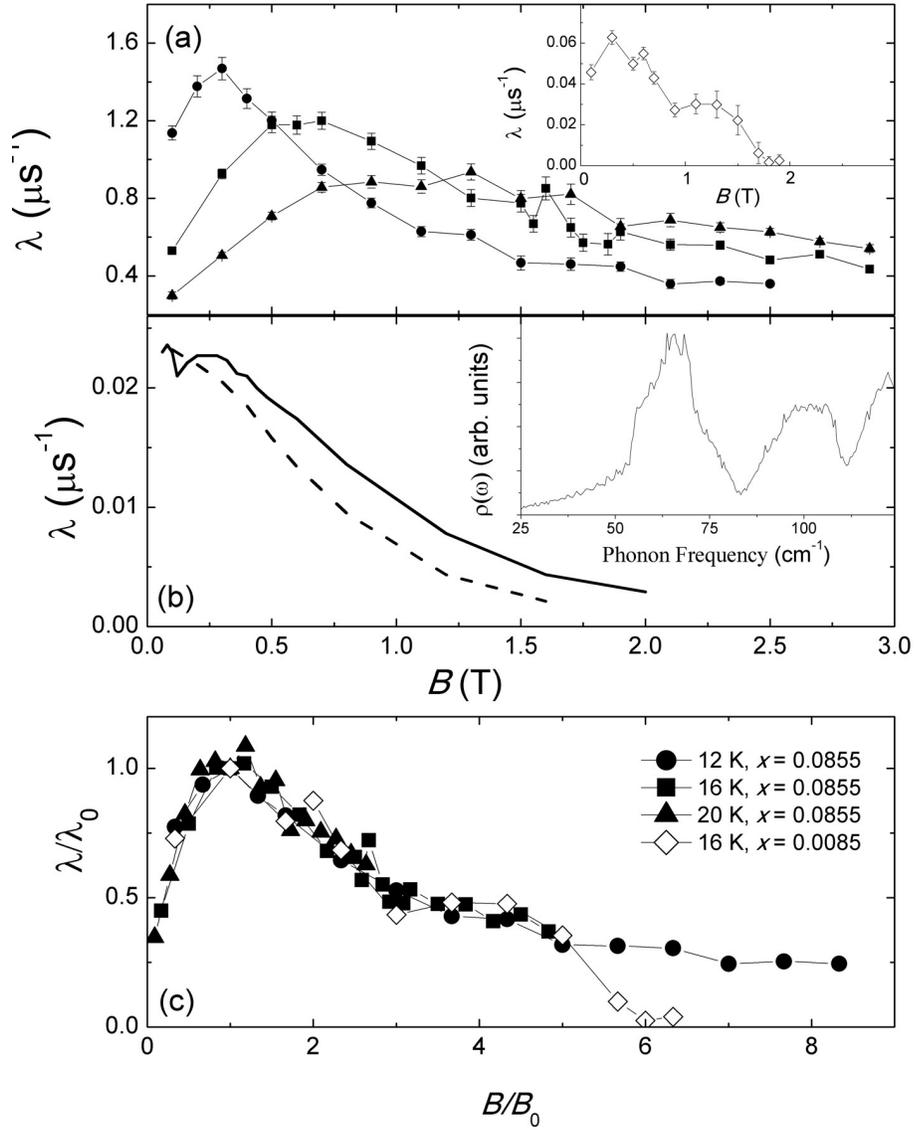

33